# *MeFiT*: *Me*rging and *Fi*ltering *T*ool for Illumina Paired-End Reads for 16S rRNA Amplicon Sequencing

Hardik I. Parikh[1] (parikhhi@vcu.edu)

Vishal N. Koparde[2] (vnkoparde@vcu.edu)

Steven P. Bradley[1] (bradleysp@vcu.edu)

Gregory A. Buck[1,2] (gabuck@vcu.edu)

Nihar U. Sheth[2*] (nsheth@vcu.edu)
*Corresponding Author

[1]Department of Microbiology and Immunology, Virginia Commonwealth University, Richmond, Virginia USA

[2]Center for the Study of Biological Complexity, Virginia Commonwealth University, Richmond, Virginia USA

**Abstract**

Recent advances in next-generation sequencing have revolutionized genomic research. 16S rRNA amplicon sequencing using paired-end sequencing on the MiSeq platform from Illumina, Inc., is being used to characterize the composition and dynamics of extremely complex/diverse microbial communities. For this analysis on the Illumina platform, merging and quality filtering of paired-end reads are essential first steps in data analysis to ensure the accuracy and reliability of downstream analysis. We have developed the Merging and Filtering Tool (MeFiT) to combine these pre-processing steps into one simple, intuitive pipeline. MeFiT provides an open-source solution that permits users to merge and filter paired end illumina reads based on user-selected quality parameters. The tool has been implemented in python and the source-code is freely available at https://github.com/nisheth/MeFiT.

**Background**

Since its inception, next-generation sequencing (NGS) techniques have transformed the way scientists extract multifaceted biological information from complex systems, fostering research in the fields of human disease, environmental science, evolutionary science, etc.[1] The exploration of bacterial ecology by sequencing the small-subunit 16S rRNA gene, or more commonly, portions thereof, is being used as a gold-standard technique for various human/environmental microbiome studies.[2-6] Thus, the predictable conservation and variability of different portions of the prokaryotic 16S rRNA gene have been exploited to provide high resolution identification and quantification of the bacterial source.[6] This strategy for phylogenetic analysis and quantification has proven to be significantly more efficient than more traditional cloning and sequencing or RT-PCR based approaches. These massively parallel, cost-effective, and high-throughput sequencing technologies can now produce up to 15Gb of genomic data in one run[7]. The sheer volume and complexity of data generated by these NGS systems necessitates the development of bioinformatics tools that streamline the downstream analyses.

The Illumina platforms provide paired-end sequencing, in which a DNA sequence is read from both ends up to a specified read length. Depending on the read length selected, currently up to 300 bases, a target DNA segment that is longer than the sum of the forward and reverse reads would result in a gap of missing sequence between them, and a shorter target segment will result in an overlap between the reads. Moreover, since quality tends to degrade towards the ends of the reads, reliable merging of overlapping paired-end reads can results in a combined DNA sequence that might permit bioinformatics correction of these 3'-end sequencing errors and yield higher quality sequence output.

In paired-end sequencing, accurate merging of the forward and reverse reads is a crucial first step that affects the results of a plethora of downstream analyses, especially, including but not limited to microbial taxonomic profiling. Various tools; *e.g.,* SHERA,[8] FLASH,[9] PANDAseq,[10] COPE,[11] and others, have been developed for merging paired end data. These tools generally apply sequence alignments to attempt to identify the best overlap between the paired-end reads. Thus, they resolve mismatching bases by considering their quality scores, with the higher quality base simply replacing one with poorer quality. Recently, a new method was proposed – CASPER, context-aware scheme for paired-end reads.[12] CASPER uses the traditional quality-score based method to resolve mismatches, except when the difference in quality scores is not significant, relies on k-mer-based contexts surrounding the mismatch to make the decision.

Another important aspect of sequence data that significantly impacts downstream analysis is read quality. Sequences of poor quality (base calling errors, small insertions/deletions) need to be identified and removed prior to analysis to minimize sources of false positivity. Each NGS platform provides some quality control measures, and others have been developed for filtering of these sequencing artifacts.[13-18] Most pipelines follow a QC protocol that filters reads based on their Phred, or other equivalent quality (Q) scores. Thus, reads of a specified length with greater than or equal to a specified, and often quite arbitrary average quality score are labeled as high-quality reads, and others with lower average quality scores are labeled low quality and removed from further analysis.[18] However, Phred scores are logarithmically related to the error probability of base-calling, which means that an average Q-score of the read is not necessarily a good indicator of the expected accuracy/error within it.[19] For example, consider two reads of length 100 nucleotides – read1 with 90 nt with Q-score of 40 and 10 nt with Q-score of 2 (average Q score ~36) , and read2 with 100 nt with Q-scores of 25 (average Q score ~25)). However, the number of miscalls or maximum expected error (MEE) in read1 will be ~6; *i.e.* 10 bases each with a probability of being incorrect of 0.63. In contrast, the MEE of read2 is 0.316 (100 * 0.00316). Thus, despite its lower average Q-score, read2 has lower error probability, likely has many fewer sequence errors, and is for most purposes a higher quality read than read1.

The focus of research in our lab has been on applying high-throughput technologies to characterize the genome, transcriptome and proteome of the human microbiome and of the host-microbial interactions. We are leading bioinformatics analysis on multiple human microbiome projects including the Vaginal Human Microbiome Project (VaHMP),[20-21] the Multi-Omics Microbiome Study: Pregnancy Initiatives (MOMS-PI)[22], oral and gut microbiome in neonates.[23] Each project involves the development of specialized computational methodologies and tools for analysis of large data sets generated by NGS sequencing on Illumina platforms. The accuracy of 16S rRNA microbial profiling is highly dependent on the accuracy of pre-processing of sequencing data. Thus, we have developed a strategy entitled *MeFiT* (***Me***rging *and **Fi**ltering **T**ool*) that efficiently merges overlapping paired-end sequence reads from the Illumina MiSeq™ sequencing platform and quality filters them using the MEE measure outlined above. MeFiT invokes a version of CASPER[12] for merging paired-end reads and extends it by including careful quality-filtering. We provide users the option to quality filter the reads using the traditional average Q-score metric or using a maximum expected error cut-off threshold. We have also optimized MeFiT with the additional functionality of appending non-overlapping paired end reads; *i.e.,* paired end reads from amplicons of greater than 600 bases which do not overlap are not discarded and are used to accurately determine the taxon from which they were derived.

Most life science researchers lack an extensive IT-infrastructure to support data analysis of the scale required for projects employing high throughput NGS technologies. These investigators often cobble together a variety of software packages and pipelines to perform their analyses, often leading to a lack of quality control and computational efficiency. Herein, we describe our efforts to develop a *fast and accurate* in-house decision support tool for use as a first pre-processing step to obtain clean high-quality data for several ongoing projects generating microbiome profiles from targeting 16S rRNA sequencing using paired end sequencing technology provided by the Illumina MiSeq platform. The MeFiT pipeline combines the essential first steps in analysis of paired-end sequencing data; *i.e.,* merging of overlapping reads and quality filtering of the reads. It is freely available (https://github.com/nisheth/MeFiT) and can be easily and intuitively used by other groups dealing with similar challenges.

**Implementation**

*The MeFiT Pipeline*

The MeFiT pipeline performs following two steps: (1) merging of overlapping paired-end sequences, and (2) filtering data for quality (Figure 1). The pipeline accepts the forward and reverse reads files in the FastQ format and automates the merging and quality filtering of overlapping and non-overlapping (optional) reads. It also generates an extensive report on the quality and merging statistics.

*Merging of Paired-End Reads*: MeFiT invokes a version of the CASPER algorithm (see REF 7) to merging the forward and reverse reads generated in paired-end sequencing. Briefly, the first step of the CASPER algorithm is to identify the best possible overlap region, with the least number of mismatching bases, between the forward and reverse reads. Any mismatches are then resolved by relying on the difference in their quality scores, with a lower quality base being replaced by the one with the higher score. Where the difference is not significant, CASPER makes partial decisions on *k-mer*-based contexts around the mismatch to make a final decision for resolving the mismatch. The user, as needed, can modify the default CASPER parameters implemented into MeFit.

*Non-Overlapping Reads:* As outlined above, 2 X 300 base PE reads will not overlap if the target amplicon is greater than 600 base pairs. Although targeted 16 S amplicons are generally selected to be smaller than 600 base pairs, some bacterial taxa, including unknown taxa, occasionally exhibit target sequences exceeding the expected size. Most existing analysis packages tend to discard these sequences, resulting in the gross underrepresentation of some taxa. MeFiT permits the user to specify to retain (default: discard) non-overlapping reads by linking the forward and reverse reads with a user-specified patch of *Ns*. MeFiT reverse complements the reverse read and appends it to the 3' end of the forward linked by a string of *Ns* (default: 15) with a assigned Phred quality scores of 2, which are subsequently ignored during the filtering step and calculation of quality statistics. Thus, taxa with unexpectedly large target amplicons are retained in the analysis.

*Quality Filtering.* As outlined briefly above, we were unsatisfied with the use of a simple average Q score to identify good and bad reads. Thus, the program permits the user to select either the average read Phred-quality score or Q score, or the maximum expected error (MEE) as a percent of read length (*meep* score).[24] The maximum expected error (MEE) within a read is the sum of error probabilities of each base, given by –

$$P = 10^{-\frac{Q}{10}}$$

$$readMEE = \sum P_i$$

where, *P* is the error probability and *Q* is the Phred-quality score (represented as ASCII character) of the $i^{th}$ base of a read. The *meep*-score is then computed from the MEE of a read as the percentage of its read length. Thus, a *meep* cut-off of 1% permits a maximum error of 1 in a read of length 100 nt.

$$meep = (readMEE * 100) / read\ length$$

**Results and Discussion**

With the advent and availability of Illumina's MiSeq system, capable of generating 2 X 300 base paired-end reads and up to 15Gb per run of 600 base-pair sequence information, it has become a go-to platform for targeted gene sequencing, metagenomics, small genome sequencing, etc.. Accurate merging and quality control of raw MiSeq data is essential before any downstream analysis. Inaccurate merging or the presence of low-quality reads (unreliable base-calls) result in an increase in the false negatives and false positives in subsequent analyses. The MeFiT pipeline was developed to perform the two prerequisite steps – merging and quality filtering of the paired end reads, to obtain high-quality data prior to subsequent analyses.

As the first step of the pipeline, MeFiT uses CASPER to merge the forward and reverse reads. As previously reported, CASPER, with its novel mismatch-resolving algorithm, showed a high level of accuracy when compared to COPE, FLASH and PANDAseq.[12] Although CASPER is not the fastest approach computationally, as expected from a k-mer based approach, the higher accuracy and robustness of the software compensates for this weakness.

One of the drawbacks of most current merging tools is that paired end reads that do not merge properly are not retained, possibly leading to the erroneous elimination of these reads from downstream analysis. Also, the target DNA fragment may be longer than the sequenced read length resulting in a gap between the forward and reverse reads. For example, in 16 S rRNA-based taxonomic profiling, a targeted region of a known (or unknown) bacterial taxon may exceed the 600 base length of the 2 X 300 base paired end sequencing of the MiSeq platform. Such forward and reverse reads will not overlap, and discarding them as most analysis protocols do results in the loss of valuable information. Considering a typical paired-end sequencing experiment of 2 x 300 bp reads, one can sequence the 16S hypervariable regions up to ~540bp in length (with a 20bp forward and reverse primers on average and a minimum of 20bp overlapping region). Our analysis of the lengths of hypervariable regions (V1–V3, V2–V4,

V3–V5, V6–V9) of sequences in the SILVA database (16S, SSU Ref NR, v119)[25], extracted using V-Xtractor[26], showed that a significant fraction of taxa have hypervariable regions that would likely result in amplicons of greater than 540 bp. The most impacted taxa are the Firmicutes and Proteobacteria which are abundant in gut microbiome samples (Table S1). A similar investigation of the STIRRUPS database, which targets largely taxa of relevance to the female human urogenital tract, showed that ~1% of bacteria have V1–V3 regions in excess of 580bp when the primer sequences are included.[27] We believe that retaining such non-overlapping high-quality data will improve the accuracy of downstream phylogeny and taxonomy analysis. The MeFiT pipeline addresses the above-mentioned shortcoming of other tools and provides users with the flexibility to append forward/reverse reads where no overlap was identified by CASPER.

The merged reads (overlapping and non-overlapping) are subjected to a quality control filter. Mefit invokes two alternative quality assessment metrics: the average Q-score, or the *meep*-score. Reads that pass the user-defined filtering criteria (average Q-score above threshold, or *meep*-score below threshold) are saved as High Quality reads in FastQ format for further downstream analysis. Finally, MeFiT generates a detailed quality-statistics output for each processed sample that provides the total number of reads, the number of overlapping reads, the number of non-overlapping reads, the average read length, the average Q-score, the average meep-score in reads, the number of high-quality reads, the average read length in high-quality reads, the average Q-score in high-quality reads, the average meep-score in high-quality reads, and the percent of overlapping reads in high-quality reads. Table S2 shows the detailed statistics report for 4 samples (randomly selected in-house samples) processed through the *MeFiT* pipeline. Filtering reads for quality using an average Q-score cutoff (threshold: 20) retains > 99% reads for each sample, in contrast, filtering for quality using *meep* cutoff (threshold: 1) only results in 83% - 90% high-quality reads. This clearly indicates that quality-filtering using average Q-scores result in reads being labeled high-quality in spite of having higher error probabilities.

To demonstrate the utility of MeFiT, we generated a 2 x 300 bp simulated dataset from V1-V3 region of the 16S gene of six reference species: *Lactobacillus crispatus, L. iners, Prevotella bivia,* and *Gardnerella vaginalis,* which have V1-V3 amplicons varying from 471-519 bp; and *Clostridium josui* and *Campylobacter rectus* which have amplicons of over 595 bp. The simulated dataset, generated using Grinder[28], consisted of 120,000 paired-end reads of 2 x 300 read length configurations, 20,000 from each of the six species. Sequencing errors were introduced using a modified 4$^{th}$ degree polynome Illumina error model 3e-3 + 1.8e-9 *i^4.[29] MeFiT was run on this dataset, saving the non-overlapping reads appending them with a patch of 15 *N*s. The merged reads were filtered using a meep cut-off of 1. As seen in Table S3, paired-end reads simulated from species with shorter V1-V3 regions (less than 540bp) result in overlapping reads, while those simulated from *C. josui* and *C. rectus* are non-overlapping and the standard applications of most analyses would simply discard the reads from the two taxa with longer V1-V3 regions, resulting in a biased community profile. OTU clustering or taxonomic classification, on a set of only high-quality overlapping reads will result in a community profile composed of only the four species with shorter V1-V3 amplicons. However, accurate identification and community abundance profiles are obtained if high quality non-overlapping reads are retained (Figure S1).

Some of the common bioinformatics tools for microbiome analysis; including mothur[30,31], QIIME[32] and USEARCH,[33] have their own implementations for merging paired-end reads. The standard operating procedure for analyzing 16S rRNA sequences generated using Illumina's MiSeq platform, demonstrated by mothur,[31] suggests using the 'make.contigs' command for preprocessing reads, and subsequently discards reads that cannot be assembled into contigs and does not provide a quality-filtration step after merging. Similarly, QIIME and USEARCH have implementations that merge and quality filter overlapping paired-end reads. However, these steps are not a part of the standard analysis pipelines, resulting in the need for users to incorporate these preprocessing steps in their custom pipelines to obtain high-quality data for analysis. Our studies suggest that CASPER outperforms these more standard approaches in terms of sensitivity and specificity. Thus, we have developed a facile pipeline that takes sample-specific raw reads as input and provides a set of high-quality reads that can be directly fed into softwares/tools to perform OTU analysis, taxonomic classification, functional profiling and other microbiome analysis.

## Availability and Requirements

Project name: MeFiT

Project home page: https://github.com/nisheth/MeFiT

Operating system(s): UNIX-based platform

Programming language: Python v2.7

Other requirements: CASPER (http://best.snu.ac.kr/casper)
numpy (http://www.numpy.org/)
HTSeq(http://www-huber.embl.de/users/anders/HTSeq/doc/install.html).

License: Apache 2.0

Any restrictions to use by non-academics: Apache 2.0

## Conclusions

The MeFiT pipeline combines open-source merging tool CASPER to a quality filtering step that automates the primary steps of analyzing overlapping paired-end sequencing data. Our pipeline reduces chances of human error due to minimal input requirements. Due to its simplified implementation and configuration, it can be easily incorporated into other analyses pipelines.

## Acknowledgements

This work was partially funded by the NIH Common Fund Human Microbiome Project (HMP) program through grant 8U54HD080784 to G Buck, J Strauss, and K Jefferson. We gratefully acknowledge the technical and philosophical assistance and advice of our colleagues MG Serrano, NC Asmussen, SW Norris and other members of the Vaginal Microbiome Consortium (VMC) at VCU. We especially thank K Hendricks-Munoz and P Xu for providing the datasets for development and testing purposes.

**Figures:**

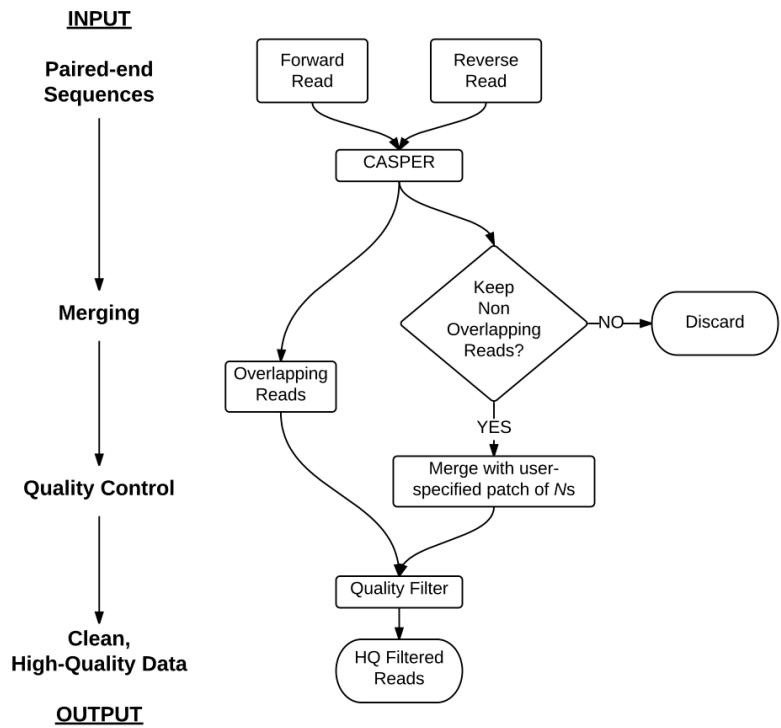

Figure 1: The *MeFiT* pipeline

**Supplementary Information:**

**Table S1**: Phylum-level distribution of sequences in SILVA rRNA database (16S, SSU Ref NR, v119) with hyper-variable regions (extracted using V-Xtractor[27]) greater than/equal to 540bp.

| Phylum | 16S Region | | | |
| --- | --- | --- | --- | --- |
| | V1-V3 | V3-V5 | V4-V6 | V6-V9 |
| Acidobacteria | 0 | 1 | 1 | 1 |
| Actinobacteria | 2 | 3 | 4 | 0 |
| Archaeplastida | 0 | 0 | 0 | 5 |
| Armatimonadetes | 0 | 0 | 0 | 1 |
| BD1-5 | 1 | 0 | 0 | 0 |
| Bacteroidetes | 1 | 11 | 16 | 2 |
| Candidate division BRC1 | 0 | 1 | 1 | 0 |
| Candidate division JS1 | 4 | 0 | 0 | 0 |
| Candidate division OD1 | 4 | 2 | 1 | 4 |
| Candidate division OP11 | 0 | 4 | 19 | 0 |
| Candidate division OP8 | 0 | 0 | 0 | 1 |
| Candidate division SR1 | 0 | 0 | 1 | 0 |
| Candidate division TM7 | 3 | 0 | 1 | 0 |
| Candidate division WS3 | 0 | 1 | 1 | 0 |
| Candidate division WS6 | 0 | 0 | 1 | 0 |
| Chloroflexi | 1 | 0 | 0 | 0 |
| Crenarchaeota | 0 | 0 | 5 | 0 |
| Cyanobacteria | 0 | 1 | 7 | 6 |
| Deferribacteres | 0 | 1 | 0 | 2 |
| Deinococcus-Thermus | 0 | 0 | 1 | 2 |
| Euryarchaeota | 0 | 0 | 33 | 1 |
| Firmicutes | 87 | 32 | 46 | 186 |
| Fusobacteria | 0 | 0 | 1 | 0 |
| Gemmatimonadetes | 0 | 1 | 0 | 0 |
| Nitrospirae | 6 | 0 | 2 | 2 |
| Opisthokonta | 0 | 0 | 0 | 13 |
| Planctomycetes | 0 | 1 | 1 | 2 |
| Proteobacteria | 57 | 66 | 97 | 375 |
| SAR | 0 | 0 | 0 | 2 |
| Spirochaetae | 1 | 0 | 1 | 1 |
| Synergistetes | 0 | 0 | 0 | 1 |
| Thaumarchaeota | 0 | 1 | 5 | 0 |
| Verrucomicrobia | 0 | 0 | 1 | 0 |
| WCHB1-60 | 1 | 0 | 0 | 0 |

**Table S2**: MeFiT sample statistics

| SampleID | Total Reads | Overlapping (%) | Non Overlapping (%) | Quality Filtering Method | Threshold | Avg Read Length | Avg Quality | Avg meep | HQ Reads (%) | HQ - % Overlapping | HQ - Avg Read Length | HQ - Avg Quality | HQ - Avg meep |
|---|---|---|---|---|---|---|---|---|---|---|---|---|---|
| Sample 1 | 46542 | 43102 (92.61%) | 3440 (7.39%) | meep | 1 | 557.55 | 33.93 | 0.63 | 38834 (83.44%) | 99.86 | 553.97 | 35.16 | 0.29 |
| Sample 2 | 46113 | 43140 (93.55%) | 2973 (6.45%) | meep | 1 | 529.52 | 34.14 | 0.57 | 39440 (85.53%) | 99.95 | 524.21 | 35.33 | 0.24 |
| Sample 3 | 35269 | 33719 (95.61%) | 1550 (4.39%) | meep | 1 | 519.6 | 34.77 | 0.42 | 31871 (90.37%) | 99.99 | 515.48 | 35.64 | 0.19 |
| Sample 4 | 26741 | 25179 (94.16%) | 1562 (5.84%) | meep | 1 | 520.65 | 34.55 | 0.48 | 23696 (88.61%) | 99.97 | 515.22 | 35.61 | 0.19 |
| Sample 1 | 46542 | 43102 (92.61%) | 3440 (7.39%) | avgq | 20 | 557.55 | 33.93 | 0.63 | 46373 (99.64%) | 92.94 | 557.38 | 22.99 | 0.61 |
| Sample 2 | 46113 | 43140 (93.55%) | 2973 (6.45%) | avgq | 20 | 529.52 | 34.14 | 0.57 | 45905 (99.55%) | 93.97 | 529.19 | 34.21 | 0.55 |
| Sample 3 | 35269 | 33719 (95.61%) | 1550 (4.39%) | avgq | 20 | 519.6 | 34.77 | 0.42 | 35138 (99.63%) | 95.96 | 519.28 | 34.83 | 0.40 |
| Sample 4 | 26741 | 25179 (94.16%) | 1562 (5.84%) | avgq | 20 | 520.65 | 34.55 | 0.48 | 26617 (99.54%) | 94.59 | 520.27 | 34.63 | 0.46 |

**Table S3:** MeFiT results for simulated dataset

| Organism | V1-V3 length | Total Reads | Overlapping (%) | Non Overlapping (%) | Avg Read Length | Avg meep |
|---|---|---|---|---|---|---|
| *Lactobacillus iners* | 519 | 20000 | 19953 (99.77%) | 47 (0.23 %) | 519.23 | 0.1 |
| *Lactobacillus crispatus* | 512 | 20000 | 19991 (99.95%) | 9 (0.04 %) | 512.05 | 0.09 |
| *Prevotella bivia* | 490 | 20000 | 20000 (100 %) | 0 (0 %) | 490 | 0.06 |
| *Gardnerella vaginalis* | 471 | 20000 | 20000 (100 %) | 0 (0 %) | 471 | 0.05 |
| *Clostridium josui* | 595 | 20000 | 27 (0.14 %) | 19973 (99.86 %) | 614.97 | 0.3 |
| *Campylobacter rectus* | 651 | 20000 | 2 (0.01 %) | 19998 (99.99 %) | 615 | 0.3 |

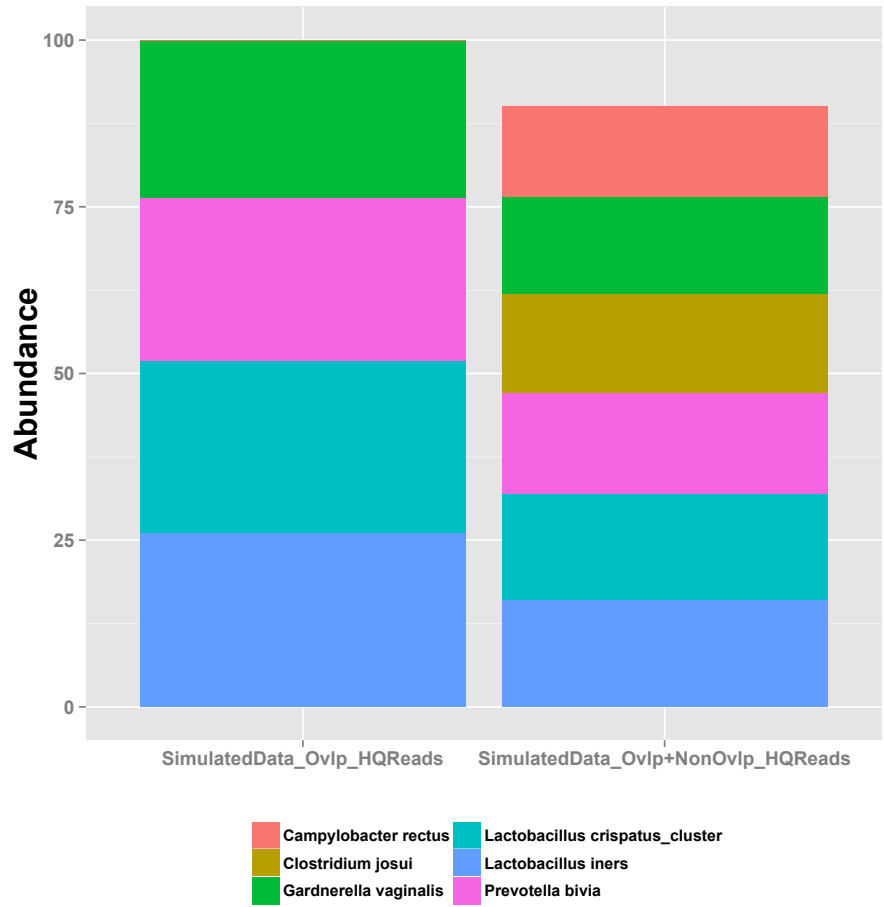

**Figure S1:** Species-level classification of simulated dataset using STIRRUPS[27]. Classifi[...] on a set of only overlapping high-quality reads result in community composed of four sp[...] with shorter V1-V3 amplicons (*G. vaginalis, L. crispatus, L. iners, P. bivia*). However, inc[...] both overlapping and non-overlapping high-quality reads result in the identification of com[...] comprising of all six species (in addition, *C. rectus and C. josui*), with more accurate re[...] abundances.